\documentclass[12pt,a4paper]{article}
\usepackage{epsfig,amsmath,amssymb,subfigure,placeins}
\usepackage[colorlinks=true,linktocpage=true,linkcolor=blue,citecolor=blue]{hyperref}
\usepackage{color}
\newcommand{\jpsi}{J / \psi}

\newcommand{\old}[1]{}

\newcommand{\be}{\begin{equation}}
\newcommand{\ee}{\end{equation}}
\newcommand{\ba}{\begin{eqnarray}}
\newcommand{\ea}{\end{eqnarray}}
\newcommand{\bi}{\begin{itemize}}
\newcommand{\ei}{\end{itemize}}

\begin{document}
\begin{flushright}
{\normalsize
}
\end{flushright}
\vskip 0.1in
\begin{center}
{\large{\bf Dissociation of quarkonium in an anisotropic hot QCD medium}}
\end{center}
\vskip 0.1in
\begin{center}
Lata Thakur$^a$,
Najmul Haque$^b$, 
Uttam Kakade$^a$,
and Binoy Krishna Patra$^a$\footnote{e-mail:binoyfph@iitr.ernet.in}

\end{center}
{\small {\it $^a$Department of Physics, Indian Institute of 
Technology Roorkee, India, 247 667}\\
 {\it $^b$Theory Division, Saha Institute of Nuclear Physics, 1/AF Bidhanagar, Kolkata-700064.}}
\vskip 0.01in
\addtolength{\baselineskip}{0.4\baselineskip} 


\begin{abstract}
We have studied the properties of quarkonium states in the
presence of momentum anisotropy by correcting the full Cornell potential
through the hard-loop resumed gluon propagator. The
in-medium modification to the potential causes less screening, so quarkonium 
states become tightly bound than in 
isotropic medium. In addition, the anisotropy in the momentum space introduces 
a characteristic angular dependence in the potential and as a result
the quark pairs aligned in the direction of 
anisotropy are bound stronger than those of perpendicular alignment. 
Since the weak anisotropy represents a perturbation to the
(isotropic) spherical potential, we use the quantum mechanical perturbation
theory to obtain the first-order correction due to the 
small anisotropic contribution to the energy eigen values of 
spherically-symmetric potential. 
\end{abstract}


\vspace{1mm}

\section{Introduction}

Ultrarelativistic heavy-ion experiments have shown very rich physics 
which cannot be interpreted by mere extrapolation from elementary 
nucleon-nucleon collisions to nucleus-nucleus collisions. This is evidenced from
the suppression of high transverse momentum region of hadron spectra up to 
a factor of 5 relative to nucleon-nucleon collisions,
which is an indication for strong absorption of high-energy partons traversing 
the medium~\cite{Wang}. Also the inclusive production of charm quark bound 
states are suppressed by a factor of 3-5 at both Super 
Proton Synchrotron (SPS) and Relativistic Heavy Ion Collider (RHIC) experiments 
which hints their dissolution in the medium~\cite{Rapp,Kluberg}.
These accumulated evidences indicate that a new form of matter
has been produced in the heavy-ion collision experiments, called quark-gluon 
plasma (QGP). Among different experimental 
observations which may be served as the signals for the QGP formation, 
quarkonium suppression has been proposed long time ago as a clear probe of 
the QGP formation in the collider experiments~\cite{McLer,adam05}. 
First, in a pioneering work by Matsui and Satz~\cite{matsui} and 
later in a follow up quantitative calculation~\cite{Karsch}, it
was shown that the suppression of $J/\psi$ yields could be explained
by the (color) screening of the potential between a heavy quark and anti-quark 
by the surrounding deconfined light quarks and gluons. 
Theoretically, one can study the quarkonium states
by using the effective field theories. Since the mass of heavy quark, $m_Q$
is much larger than the intrinsic scale in the theory of 
Quantum Chromodynamics (QCD), $\Lambda_{\rm{QCD}}$, the 
heavy quark and anti-quark are expected to move slowly with 
a relative velocity $v \ll 1$ and results in the non-relativistic version of  
QCD (NRQCD) \cite{nrqcd1,nrqcd2}. However, NRQCD
does not fully exploit the smallness of $v$ which further gives rise 
another effective theory, known as potential non-relativistic QCD (pNRQCD)~\cite{pNRQCD1,pNRQCD2}
by integrating out the momentum scale.

The heavy quark pairs formed
in relativistic nuclear collisions develop into the physical resonances 
and traverse the plasma and then hot hadronic matter
before decaying into dilepton pairs. Even before the resonance is formed
it may be absorbed by the nucleons streaming past it~\cite{GH} and by the
time the resonance is formed, the screening of the color force
in the plasma may inhibit the formation of bound states.
The resonance(s) could also be dissociated either
by hard gluons~\cite{xu,gdiss1,gdiss2,gdiss3,gdiss4} or by comoving
hot hadrons~\cite{vogt}. In order to disentangle these sequential
effects~\cite{dks}, we must
know how the properties of quarkonium states change in medium.
The basic tools of phenomenological approach to study the properties of 
quarkonium states are potential models where the possible relativistic effects 
for excited states of charmonium may also be incorporated there.
At zero temperature, the potential model
has made great success. At finite temperature, 
the essence of the potential model in the context of deconfinement is to use a
finite temperature extension of the potential. Quantitative understanding
of the bound state properties needs the exact potential at finite temperature 
which, in principle, should be derived directly from QCD, like the Cornell
potential at zero temperature has been derived from pNRQCD from the 
zeroth-order matching coefficient. Such derivations
at finite temperature for weakly-coupled plasma have been recently come up in the 
literature~\cite{Brambilla05,Brambilla08} but they
are, however, plagued by the existence of temperature-driven 
hard as well as soft scales, $T$, $gT$, $g^2 T$, respectively.
Due to these difficulties in finite temperature extension in effective
field theories, the lattice-based potentials become popular. 
However, neither the free energy nor the internal energy 
can be directly used as the potential. 
In fact, what kind of screened potentials should be used 
in the Schr\"odinger equation which describes well the bound states at finite 
temperature are still an open question. 
However, recently more involved calculations of quarkonium spectral 
functions and meson current correlators obtained from potential 
models have been performed and compared to 
first-principle QCD calculations performed numerically on 
lattices~\cite{Asak04,Dat04,Hatsuda,Jakovac,Aarts11}.
In addition to the uncertainties of the correct form of the
finite-temperature potential, there are also arbitrariness of 
the criteria of dissociation. Besides the binding energy 
for a particular potential as an criterion of dissociation, the decay width 
is another important quantity 
to determine the dissociation of the bound states. 
The calculation based on a real-valued potential model does not include the 
true width of a state. By performing 
an analytical continuation of the Euclidean Wilson loop to Minkowski space, 
the potential has an imaginary part due to Landau damping which clearly 
broadens the peak~\cite{Brambilla08,Laine1} and facilitates the early
dissociation of the bound states.

In the RHIC or LHC era (small $\mu_{B}$), recent lattice studies have 
confirmed that the transition from
nuclear matter to QGP is not a phase transition, rather a
crossover\cite{phaseT}.
The large distance property of the heavy quark
interaction is important for our understanding of the bulk properties of
the QCD plasma phase, {\it e.g.} the screening property of the quark gluon
plasma \cite{Kar62+70}, the equation of state \cite{Dum05,Dum2} etc. 
In these studies, deviations from perturbative 
calculations and the ideal gas behavior are found
beyond the deconfinement temperature. 
It is then reasonable to assume that the string-tension
does not vanish abruptly at the deconfinement point~\cite{string1,string2,string3}, so
one should study its effects on heavy quark potential even above $T_c$.
This issue, usually overlooked in the
literature where only a screened Coulomb potential was
assumed above $T_c$ and the linear/string term was assumed zero,
was certainly worth investigation.
Recently a heavy quark potential at finite
temperature was derived by correcting the full Cornell potential,
not its Coulomb part alone,
with a dielectric function encoding the effects of the deconfined 
medium~\cite{prc_vineet}. This was found to have an additional
long range Coulomb term, in addition to the conventional Yukawa term. 
In the short distance limit, the potential is reduced to vacuum potential, 
{\it i.e.}, $Q \bar Q$ pair does not see the medium, giving rise the duality
between $V(r,T=0)$ and $V(0,T)$. On the
other hand, in the large distance limit (where the screening
occurs), potential is reduced to a long-range Coulomb potential with
a dynamically screened-color charge. Thereafter the binding energies and 
dissociation temperatures
of the ground and the lowest-lying states of charmonium and bottomonium
spectra have been determined~\cite{prc_vineet,chic_vineet} which 
matches with the finding of recent works based on
potential models~\cite{cabrera,mocsyprl,digal} with the Debye mass
extracted from the lattice free energy. However, when the Debye mass 
in leading-order was used, the results~\cite{prc_vineet}  
match with the lattice correlator studies or 
with the stronger (lattice) binding potential, i.e. internal energy~\cite{wong1,wong2}.
In a way, their findings address the reason of arbitrariness of the results 
on dissociation temperatures~\cite{Satz} and ensues a basic question about the nature of dissociation
of quarkonium in a hot QCD medium.

However, all the works described above was limited to an isotropic 
medium but the partonic system generated in an ultra-relativistic 
heavy-ion collisions cannot be homogeneous and isotropic
because at the very early time of collision, asymptotic weak-coupling enhances
the longitudinal expansion substantially than the radial expansion so
the system becomes colder in the longitudinal direction than in the 
transverse direction. As a result, an anisotropy in the momentum space sets in and 
causes the parton system produced
unstable with respect to the chromomagnetic plasma modes~\cite{Romatschke} 
which facilitate to isotropize the system~\cite{Arnold05,Mrowczy93}.
In our work, we restricted ourselves to a {\em weakly anisotropic medium}
because by the time ($t_F$ =$\gamma \tau_F$, $\tau_F$ is the quarkonium 
formation time in its rest frame) quarkonia are formed in the plasma, the 
plasma becomes almost equilibrated. Motivated with this preamble on
the anisotropy generated in the very stage of collision, we wish to
investigate the effect of weak anisotropy on the heavy-quark potential
and subsequently on the dissociation of quarkonia states in an anisotropic
medium.

Our work is organized as follows. In Section~\ref{poten}, we will discuss the 
medium modifications to a heavy quark potential both in isotropic and
anisotropic medium. In Sec.\ref{iso}, we start with the in-medium modification
to the heavy quark potential in an isotropic medium and then 
extend it to a medium which exhibits a local anisotropy in momentum 
space in Sec.\ref{aniso}. 
To do that, we first obtain the self-energy tensor for an 
anisotropic medium to obtain the hard-loop resumed 
gluon propagator. Thereafter the dielectric permittivity be 
obtained
in terms of retarded gluon propagator and its Fourier transform at vanishing frequency 
gives the desired non-relativistic potential at finite temperature. The
potential thus obtained depends not only on the relative separation
of $Q \bar Q$ pair but also on their relative orientation
with respect to the direction of anisotropy and 
is found always deeper than in an isotropic medium.
It is found that in the weak-anisotropy limit, the correction
arising due to anisotropy to the isotropic part of the potential is small
and thus has been treated as a perturbation.
So, using the first-order perturbation theory, 
we estimate the shift in energy eigen values due to the small 
anisotropic correction 
to the energy eigen values from the spherically-symmetric
part in isotropic medium and determine their dissociation 
temperatures in Sec.3. Finally, we conclude 
in Section~\ref{conclu}.

\section{Heavy-quark effective potential}\label{poten}
\subsection{For isotropic medium {\texorpdfstring{$(\xi = 0)$}{xi=0}}}\label{iso}

Potential models are based on the assumption that the
interaction between a heavy quark and its anti-quark
can be described by a potential.
At $T=0$, the hierarchy of well separated energy-scales: 
$m_Q\gg m_Qv\gg m_Qv^2$,
allows one to systematically integrate out the different
scales and obtain the non-relativistic potential
QCD (pNRQCD) where the Cornell potential indeed shows up as the
zeroth-order matching coefficient~\cite{pNRQCD1,pNRQCD2}. Inspired 
by its success at
zero temperature, the potential model has been applied at finite
temperature, with the main assumption that medium effects can be accounted 
by a temperature-dependent potential.

Recent lattice results~\cite{phaseT} indicate the phase transition in full QCD
appears to be a crossover rather than a `true' phase transition with the 
related singularities in thermodynamic observables.
In light of the above findings, one cannot simply
ignore the effects of string tension between the quark-anti-quark pairs
beyond $T_c$. This is indeed a very important effect which needs
to be incorporated while setting up the criterion for the dissociation.
Recently this issue
has successfully been addressed for the dissociation of quarkonium in
QGP by one of us~\cite{prc_vineet,chic_vineet}
and we closely follow their work in brief.
Let us now start with a heavy quark potential (Cornell potential) at T=0 
\begin{equation}
\label{eqc}
 V(r)=-\frac{\alpha}{r}+\sigma r \quad ,
\end{equation}
where $\alpha$ and $\sigma$ are the phenomenological parameters. The former
accounts for the effective coupling between a heavy quark and its anti-quark
and the latter gives the string coupling.
The medium modification enters through the Fourier transform of
heavy quark potential as 
\begin{equation}
\label{eq1}
\tilde{V}(k)=\frac{V(k)}{\epsilon(k)} \quad ,
\end{equation}
where $V(k)$ is the Fourier transform (FT) of the Cornell potential
which requires a regularization. We regulate both terms in the
potential by multiplying with an exponential damping factor and
is switched off after the FT is evaluated. 
This can be done by assuming $r$- as distribution
($r \rightarrow$ $r \exp(-\gamma r))$. The FT of
the linear part $\sigma r\exp{(-\gamma r)}$ is
\begin{eqnarray}
\label{eq-6-3}
=-\frac{i}{k\sqrt{2\pi}}\left( \frac{2}{(\gamma-i k)^3}-\frac{2}{(\gamma+ik)^3}
\right).
\end{eqnarray}
After putting $\gamma=0$,  we obtain the FT of the linear term $\sigma r$ 
as,
\begin{equation}
\label{eq-6-4}
\tilde{(\sigma r)}=-\frac{4\sigma}{k^4\sqrt{2\pi}}
\end{equation}
and for the full Cornell potential, the FT is 
\begin{equation}
\label{vk}
{V}(k)=-\sqrt{(2/\pi)} \frac{\alpha}{k^2}-\frac{4\sigma}{\sqrt{2 \pi} k^4}.
\end{equation}
The dielectric permittivity, $\epsilon(k)$ is given in terms
of the static limit of the longitudinal part of the gluon
self-energy~\cite{schneider}. In isotropic case, it can be 
decomposed into longitudinal ($\Pi_L$) and transverse ($\Pi_T$) 
components which, in the static limit, are associated with 
the screening of electric and magnetic fields, respectively. However, in the
static limit, the transverse part
$\Pi_T (0,k\rightarrow 0,T)$ vanishes, {\em i.e.}, static magnetic 
fields are not screened.
In perturbation theory, the quantity that enters in the Fourier transform 
of the potential at finite temperature is the static limit of
the longitudinal gauge boson self-energy which
was calculated long ago~\cite{Shur78} at one loop level 
\begin{eqnarray}
\lim_{k\rightarrow 0} \Pi_L (0,k,T)=g^{2} T^{2}\left(\frac{N_c}{3}+\frac{N_f}{6}\right)
\equiv m_{{}_D}^2 \quad ,
\label{debye}
\end{eqnarray}
where $ m_{{}_D} $ is defined as the screening mass. Since the static 
limit of the 
self-energy is momentum independent, the pole of the inverse of dielectric
permittivity is simply the gauge invariant Debye 
mass $m_{{}_D} $, so it leads to an exponential damping of the
potential $V(r)\sim \exp(-m_{{}_D}r)/r$. In particular, this form of $ \Pi_L $
has the consequence that gluons screen the strong interaction, in contrast 
to the zero temperature case, over long-distance scale. If one
assumes non-perturbative effects such as the string
tension which survives even above the deconfinement point~\cite{bkp}
then the dependence of the dielectric function on
the Debye mass may get modified. 
However, we assume the same screening mass scale $m_{{}_D}$ which
emerges in the Debye screened Coulomb potential also appears in the
non-perturbative long-distance contribution due to string.
In the following section, we take over this assumption to
anisotropic medium too. However, different scales for the
Coulomb and linear pieces of the T=0 potential, rather than a single 
one, was already employed in Ref.~\cite{megiasind,megiasprd}.
Moreover, they developed a theoretical model to include non-perturbative 
effects beyond the deconfinement temperature through dimension-two gluon 
condensates to calculate the heavy quark free energy.
Interestingly, their model predicts a duality between the zero temperature 
$Q \bar Q$ potential and the quark self energy and explains the lattice 
data well.

\noindent Finally, one can define a dielectric permittivity
in one-loop by
\begin{eqnarray}
\label{eqn2}
\epsilon(k)=\left(1+\frac{\Pi_L (0,k,T)}{k^2}\right)
\equiv \left( 1+ \frac{m_{{}_D}^2}{k^2} \right)~.
\end{eqnarray}
After substituting the dielectric permittivity in the Fourier 
transform of Cornell potential (\ref{eq1}) and then evaluating its 
inverse FT, one obtains the medium modified
potential in the co-ordinate space~\cite{prc_vineet,chic_vineet}
\begin{eqnarray}
\label{defn}
V(r,T)&=&\int \frac{d^3\mathbf k}{{(2\pi)}^{3/2}}
 e^{i\mathbf{k} \cdot \mathbf{r}}~
\tilde{V}(k) \nonumber\\
&=&\left(\frac{2\sigma}{m_D}-\alpha m_D\right)\frac{\exp{(-\hat r)}}{\hat r}
-\frac{2\sigma}{m_D \hat r}+\frac{2\sigma}{m_D}-\alpha m_D
\end{eqnarray}
with the dimensionless variable $\hat{r}=m_{{}_D} r$. The constant terms are introduced  to yield 
the correct limit of $V(r,T)$ as $T\rightarrow 0$. Such terms could arise
naturally from the basic computations of real time static potential
in hot QCD~\cite{Laine2} and also from the real and imaginary time
correlators in a thermal QCD medium~\cite{beraudo}.
The medium modified potential thus obtained has an additional 
long range Coulomb term with an (reduced) effective charge, in addition
to the conventional Yukawa term. 

In the small distance limit, $r\ll 1/m_{{}_D}$, the above potential reduces to
the Cornell potential, {\it i.e.} $Q \bar Q$ does not see the medium.
On the other hand, in the screening region 
$r\gg1/m_{{}_D}$, the potential (\ref{defn}) reduces to 
\begin{eqnarray}
\label{lrp}
{V(r,T)}\sim -\frac{2\sigma}{m^2_{{}_D}r}-\alpha m_{{}_D},
\end{eqnarray}
which, apart from a constant term, looks like a Coulomb potential encountered
in hydrogen-atom problem after identifying the fine structure constant
$e^2$ with the effective charge $2 \sigma/m_{{}_D}^2$. 
The binding energies and the dissociation temperatures for 
quarkonium states can thus be determined by solving the Schr\"odinger
equation numerically either with the full potential (\ref{defn}) or 
analytically with the approximated form (\ref{lrp}). 
In addition, one can also exploit the advantage to demonstrate the flavor 
dependence of the dissociation process where the dissociation temperatures 
for 2-flavor are found to be higher than 
the 3-flavor case~\cite{prc_vineet,chic_vineet}.

\subsection{For anisotropic medium {\texorpdfstring{$(\xi \ne 0)$}{xi=0}}}\label{aniso}
\subsubsection{Dielectric permittivity tensor}
To study the perturbative potential with an anisotropic parton 
distribution, consider a hot QCD plasma which, due to expansion 
and finite (momentum) relaxation time, manifests a local anisotropy in
momentum space through the distribution function 
\begin{equation}                                        
f_{\rm{aniso}} (\mathbf{k}) = f_{\rm{iso}}\left( \sqrt{\mathbf{k}^{2} + \xi(\mathbf{k}.\mathbf{n})^{2}}  \right) \quad,
\end{equation}                                            
{\it i.e.}, $f_{\rm{aniso}} (\mathbf{k})$ is obtained from an isotropic distribution $f_{iso} (|\mathbf{k}|)$ 
by removing particles with a large momentum component along
the direction of anisotropy, $\mathbf{n}$~\cite{Romatschke}. We shall restrict 
ourselves to a plasma close to equilibrium and so that $f_{\rm{aniso}}(\mathbf{k})$ 
is either a Bose-Einstein $n_{B} (k)$ or a Fermi-Dirac
$n_{F} (k)$ distribution function. This may be true because by the 
time quarkonia have been formed in the plasma medium from the
$Q \bar Q$ pairs produced at very early stages of the collision, the system 
may not be then highly anisotropic 
rather closer to isotropic distribution. In the limit of small anisotropy,
anisotropy parameter $ \xi $ is related to the shear 
viscosity-to-entropy density ($\eta/s$) through the one-dimensional Navier 
Stokes formula by
\begin{equation}
\xi= \frac{10}{T \tau}~\frac{\eta}{s},~                 
\end{equation}                                   
where $1/\tau$ denotes the expansion rate of the fluid element.
However, the degree of momentum-space anisotropy is generically 
defined by the parameter,
\begin{equation}
\xi = \frac{\langle \mathbf{k}_{T}^{2}\rangle}{2\langle k_{L}^{2}\rangle}-1~,~
\label{anparameter}
\end{equation}
where $ {k}_{L}= \mathbf{k}.\mathbf{n} $ and ${\bf k}_T = 
\mathbf{k}-\mathbf{n}(\mathbf{k}.\mathbf{n}) $ are the components of momentum
parallel and perpendicular to the direction of 
anisotropy, $\mathbf{n}$, respectively.
The positive and negative values of $ \xi$ corresponds to the 
squeezing and the stretching of the distribution function in the direction of 
anisotropy, respectively. In the relativistic nucleus-nucleus 
collisions, $\xi$ is however, found to be positive.
The calculation of the real part of the potential at finite  anisotropy was 
first obtained in Ref. \cite{Laine2,Romatschke,Dumitru08} and  was later 
extended to calculate the imaginary 
part~\cite{Brambilla08,beraudo,Laine09,Dumitru09} which is seen as a generic
feature of the medium. 
To study the effect of anisotropy on the in-medium potential, one need 
to calculate first the self-energy in an anisotropic medium. With the specified anisotropic 
distribution function, we can compute the gluon self-energy 
analytically~\cite{Thoma62}. 
We will restrict our consideration to the spatial part of the self-energy, $\Pi^{\mu\nu} $ 
for  simplicity and the time-like components can be easily obtained 
by using the symmetry and transversality of the gluon self-energy tensor.
The spatial components of the retarded self-energy tensor reads~ \cite{Dumitru08}
\begin{equation}
\Pi^{ij}(P)= -g^{2}\int d^{3}\mathbf{k}~v^{i}\frac{\partial f(\mathbf{k})}{\partial 
k^{l}}\bigg(\delta^{jl}+\frac{v^{j}p^{l}}{ P\cdot V+i\epsilon}
\bigg) \quad,
\label{spatialselfeenrgy}
\end{equation}
where $P^{\mu}\equiv(p_0,p)$ is the four momentum of the external gluon, 
$p=|{\bf p}|=$ is the amplitude of the spatial momentum. The four-velocity,
$V^\mu$ $(1,{\bf v}={\bf k}/|\bf k|)$ is a light-like four vector
with $v=|{\bf v}|$ and the partial-derivative, 
$\partial f /\partial k^l$, in terms of new variable $\tilde{k}$, is given by 
\begin{equation}
\frac{\partial f(\mathbf{k})}{\partial k^{l}}= \frac{v^{l}+\xi(\mathbf{v} .\mathbf{n})n^{l}}
{\sqrt{1+\xi(\mathbf{v} .\mathbf{n})^{2}}}\frac{\partial f(\tilde {k}^{2})}{\partial\tilde {k}},
\end{equation}
where $\tilde{k}^{2}= k^{2}(1+\xi(\mathbf{v} .\mathbf{n})^{2})$.  After 
integrating out over the modified momentum $\tilde {k}$, the 
self-energy, $\Pi_{ij}$ is simplified into~\cite{Dumitru08}
\begin{equation}
\Pi^{ij}(P)= m_{{}_D}^{2}\int \frac{d\Omega}{4\pi}v^{i}\frac{v^{l}+\xi(\mathbf{v} .
\mathbf{n})n^{l}}{(1+\xi(\mathbf{v} .\mathbf{n})^{2})^{2}}\bigg(\delta^{jl}+\frac{v^{j}p^{l}}{P\cdot
 V+ i \epsilon}\bigg) \quad,
\end{equation}
where the square of the Debye mass is defined by
\begin{equation}
m_{{}_D}^{2}= -\frac{g^{2}}{2\pi^{2}}\int\limits_{0}^{\infty} dkk^{2}\frac{d f_{iso}
(k^{2})}{dk}.
\label{deb}
\end{equation}
Unlike in isotropic medium, the self-energy $\Pi^{\rm{ij}}$ shows an 
extra dependence on the preferred anisotropic 
direction ($\mathbf{n}$), therefore, it can no longer be decomposed 
into transverse and longitudinal parts rather it becomes a tensor 
\cite{Romatschke,Strickland,Strick04} with more basis vectors. 
So the self-energy tensor 
be decomposed into four structure functions as~\cite{Romatschke}
\begin{equation}
\Pi^{ij}= \alpha A^{ij}+ \beta B^{ij} + \gamma C^{ij} + \delta D^{ij}~,~
\end{equation}
where the coefficients $\alpha$, $\beta$, $\gamma$ and $\delta$ can be 
determined for any value of $\xi $~\cite{Romatschke}. 
Using these structure functions, one can find the gluon propagator in temporal axial gauge as~\cite{Romatschke}
\begin{eqnarray}
\Delta_{ij}(\omega,k)&=&\frac{(A_{ij}-C_{ij})}{k^{2}-\omega^{2}+\alpha}
+\frac{(k^{2}- \omega^{2}+\alpha +\gamma)B_{ij}}{(k^{2}-\omega^{2}+\alpha +\gamma)
(\beta-\omega^{2})-k^{2}{\tilde n}^{2}\delta^{2}}\nonumber\\
&&+\frac{(\beta-\omega^{2})C_{ij}-\delta D_{ij}}{(k^{2}-\omega^{2}+\alpha
 +\gamma)(\beta-\omega^{2})-k^{2}{\tilde n}^{2}\delta^{2}}
\end{eqnarray}
In order to see how the anisotropy affects the
response to static electric field, we examine the
propagator in the  static limit ($\omega\rightarrow 0$). Defining the masses
in the static limit~\cite{Romatschke}
\begin{eqnarray}
m_{\alpha}^{2}&=&\lim_{\omega \to 0}\alpha,~~
m_{\beta}^{2}=\lim_{\omega \to 0}-\frac{k^{2}}{\omega^{2}}\beta \nonumber\\
m_{\gamma}^{2}&=&\lim_{\omega \to 0}\gamma,~~
m_{\delta}^{2}=\lim_{\omega \to 0}\frac{\tilde n k^{2}}{\omega}{\rm Im} \ \delta \quad,
\label{mminus}
\end{eqnarray}
the gluon propagator becomes 
\begin{eqnarray}
\label{epsaniso}
\lim_{\omega \rightarrow 0} \Delta_{ij}(\omega,k)=-\frac{(k^{2}+m_{\alpha}^{2}+m_{\gamma}^{2})\ k_ik_j}
{\omega^2\left[(k^{2}+ m_{\alpha}^{2}+m_{\gamma}^{2})(k^{2}+m_{\beta}^{2})-
m_{\delta}^{4}\right]}~.
\end{eqnarray}
We can now factorize the denominator of the gluon propagator as 
\begin{eqnarray}
(k^{2}+ m_{\alpha}^2+m_{\gamma}^2)(k^2+m_\beta^2)- 
m_{\delta}^4=(k^2+m_+^2) (k^2+m_-^2)
\end{eqnarray}
where
\begin{eqnarray}
2 m_\pm^2=M^2 \pm \sqrt{M^4-4(m_\beta^2(m_\alpha^2+m_\gamma^2)-m_\delta^4)}\ 
,\ \ M^2=m_\alpha^2+m_\beta^2+m_\gamma^2~.
\end{eqnarray}
Thus the dielectric permittivity for an anisotropic medium in the temporal 
axial gauge can be obtained from the definition~\cite{kapusta}
\begin{eqnarray}
\epsilon^{-1}(k)=-\lim_{\omega \to 0} {\omega^2\frac{k_ik_j}{k^2}\Delta_{ij}(\omega,k)}
=\frac{k^{2}(k^{2}+m_{\alpha}^{2}+m_{\gamma}^{2})}
{(k^2+ m_+^2)(k^{2}+ m_-^2)}~,
\label{permittivity}
\end{eqnarray}
If the anisotropy is small, the pole masses (\ref{mminus}) can be simplified, 
by retaining only the linear term in $\xi$, into~\cite{Romatschke} 
\begin{eqnarray}
m_\alpha^2&=&-\frac{\xi}{6}(1+\cos 2 \beta_n)~m_{{}_D}^2\nonumber\\
m_\beta^2&=&\left( 1+\frac{\xi}{6}(3\cos 2 \beta_n-1)\right)~m_{{}_D}^2\nonumber\\
m_\gamma^2&=&\frac{\xi}{3} \sin^2\beta_n~m_{{}_D}^2\nonumber\\
m_\delta^2&=&-\xi\frac{\pi}{4}\sin \beta_n \cos\beta_n~m_{{}_D}^2
\end{eqnarray}
where $\beta_n$ is the angle between ${\bf k }$ and ${\bf n}$ .
So the explicit dependencies of $m_\pm$ on the anisotropy, in the small
$\xi$ limit, are given by
\begin{eqnarray}
m_+^{2}&=&\left(1+\frac{\xi}{6}(3\ \cos2\beta_n-1)\right)m_{{}_D}^{2},\nonumber\\
m_-^{2}&=&-\frac{\xi}{3}\ \cos2\beta_n m_{{}_{D}}^{2}.
\label{mplus}
\end{eqnarray}
In the isotropic limit, all masses become zero except
one ($m_\alpha^2 = m_\gamma^2= m_\delta^2= m_-^2=0, m_+^2=m_{{}_D}^2$),
which is the only pole in the isotropic medium (\ref{eqn2}).
\subsubsection{Medium-modification to heavy quark potential}
Once we have obtained the dielectric permittivity in anisotropic 
medium (\ref{permittivity}), we substitute it in the Fourier
transform (\ref{eq1}) and then evaluate its inverse Fourier transform
to obtain the medium modified
potential in an anisotropic medium:
\begin{eqnarray}
V(\mathbf{r},\xi,T)&=&\frac{1}{{(2\pi)}^{3/2}}
\int d^{3}\mathbf{k}~\tilde{V}(k)~e^{i\mathbf{k} \cdot \mathbf{r}}\nonumber\\
&=& -\frac{\alpha}{2\pi^{2}}\int d^{3}\mathbf{k} 
~\frac{(k^{2}+m_{\alpha}^{2}+m_{\gamma}^{2})}{(k^{2}+m_+^2)
(k^{2}+m_-^{2})}~e^{i\mathbf{k} \cdot \mathbf{r}}
\nonumber\\
&& -\frac{4\sigma}{(2\pi)^{2}}\int d^{3}\mathbf{k}~ 
 \frac{(k^{2}+m_{\alpha}^{2}+m_{\gamma}^{2})}{k^{2}(k^{2}+m_+^{2})
(k^{2}+m_-^{2})}~e^{i\mathbf{k} \cdot \mathbf{r}}
\label{modipot}
\end{eqnarray}
After substituting the pole masses, $m_+$ and $m_-$ (in the small $\xi$ limit) from (\ref{mplus}), the potential becomes
\begin{eqnarray}
V({\bf r},\xi,T)&=&\!\!\!\!-\frac{\alpha}{2\pi^{2}}\int \frac{d^{3}\bf{k}  
~e^{i\bf{k} \cdot\bf{r}}}{
k^2+m_{{}_D}^2\left(1+\frac{\xi}{6}(3\cos2\beta_n-1)\right)
} \nonumber\\
&&\!\!\!\!-\frac{4\sigma}{(2\pi)^{2}}
\int \frac{d^{3}\mathbf{k}~e^{i\mathbf{k} \cdot \mathbf{r}}}
{k^{2}\left[k^{2}+m_{{}_D}^{2}\left(1+\frac{\xi}{6}(3\cos2\beta_n-1)\right)\right]}\nonumber\\
&\equiv &\!\!\!\!V_1({\bf r},\xi,T)+V_2({\bf r},\xi,T)
\label{v1}\ ,
\end{eqnarray}
where $V_1 ({\bf r},\xi,T)$ and $V_2(\mathbf{r},\xi,T)$ are the 
medium-modified potential corresponding to short-distance Coulombic and 
long-distance string term, respectively, can be rewritten as 
\begin{eqnarray}
V_1({\bf r},\xi,T) =-\frac{\alpha}{2\pi^{2}}\int d^{3}\mathbf{k} 
e^{i\mathbf{k} \cdot \mathbf{r}}\frac{1}{\left( k^{2}+{m_{D}}^{2}\right)}\left(1+\frac{\xi}{6}~\frac{{m_{D}}^{2}}{\left( k^{2}+{m_{D}}^{2}\right)}~(3\cos2\beta_n-1)\right)^{-1}
\end{eqnarray}
Expanding the integrand in terms of the anisotropy parameter ($\xi$) and 
retaining the term linear in $\xi$ (weak-anisotropy limit, $\xi<1$),
$V_1({\bf r},\xi,T)$ can be written as
\begin{eqnarray}
V_1({\bf r},\xi,T) &=&-\frac{\alpha}{2\pi^{2}}\int d^{3}\mathbf{k} 
e^{i\mathbf{k} \cdot \mathbf{r}}\left[ \frac{1}{\left( k^{2}+{m_{D}}^{2}\right)}-\frac{\xi}{6}~\frac{{m_{D}}^{2}}{\left( k^{2}+{m_{D}}^{2}\right)^{2}}~(3\cos 2\beta_n-1)\right]\nonumber\\
&\equiv & V_1^{(1)}({ r},\xi =0,T)+ V_1^{(2)}({\bf r},\xi,T)
\end{eqnarray}
where $V_1^{(1)}({ r},\xi=0,T)$ and $V_1^{(2)}({\bf r},\xi,T)$ are 
the isotropic and anisotropic contributions due
to the medium modification of the Coulomb term, respectively.
Similarly $V_2({\bf r},\xi,T)$ can be decomposed into isotropic and anisotropic
parts :
\begin{eqnarray}
V_2({\bf r},\xi,T)=&V_2^{(1)}({r},\xi =0,T)+ V_2^{(2)}({\bf r},\xi,T)
\end{eqnarray}
where $V_2^{(1)}({r},\xi=0,T)$ and $V_2^{(2)}({\bf{r}},\xi,T)$ are the
isotropic and anisotropic contributions due
to the medium modification of the linear term, respectively.
Let us now calculate them one-by-one. 

The isotropic part, $V_1^{(1)}({ r},\xi=0,T)$ of
the Coulomb term (already calculated in (\ref{defn}))
is given by ($\hat{r}=r~m_D$)
\begin{eqnarray}
V_1^{(1)}({r}, \xi =0,T)
=-\frac{\alpha m_{{}_D}}{\hat{r}}e^{-\hat{r}} -\alpha m_{{}_D}\ ,
\label{col1}
\end{eqnarray}
and the anisotropic part, $V_1^{(2)}({\bf r},\xi,T)$ is given by 
\begin{eqnarray}
V_1^{(2)}({\bf r},\xi,T) = -\xi~\frac{\alpha~ m_{{}_D}^2}{2 \pi^2}\int 
\frac{d^3\mathbf{k} e^{i \mathbf{k} \cdot \mathbf{r}}}{(k^2+m_{{}_D}^2)^2}
~\left(\frac{2}{3}-\cos^2\beta_n\right). 
\end{eqnarray}
One immediately observes that unlike the isotropic part $V_1^{(1)}(r, \xi 
=0,T)$, momentum anisotropy ($\xi \ne 0$) causes the potential to depend
on angle, in addition to inter-particle distance (r).
Before deriving a general angular dependence we first illustrate the two 
cases which are especially interesting to grasp the effect of 
anisotropy on the heavy-quark potential.
However they will be used further to derive the general angular dependence.
First, we consider that $ \mathbf{r} $ is parallel to the 
direction of anisotropy, $\mathbf{n}$, where we have taken the 
direction of anisotropy $\mathbf{n}$ along the $z$-axis and 
the angle between the $r$ and $\bf k$ is $\theta$. 
So the anisotropic part, $ V_1^{(2)}({\bf r},\xi,T)$ for the medium
modification to the Coulomb term becomes
\begin{eqnarray}
 V_1^{(2)}({\bf r}||{\bf n},\xi,T)&=&-\xi~\frac{\alpha m_{{}_D}^2}{\pi}~
~\int \frac{d^3 \mathbf{k}~e^{i{\bf k} \cdot {\bf r}}}{(k^2+m_{{}_D}^2)^2}~ 
\left(\frac{2}{3}-\cos^2\theta\right).
\end{eqnarray}
After the angular integration, it becomes simplified as
\begin{eqnarray}
 V_1^{(2)}({\bf r}||{\bf n},\xi,T)
&=& \xi~V_1^{(1)}({ r},\xi=0,T)\left(2~\frac{e^{\hat{r}}-1}{\hat{r}^2}-
\frac{2}{\hat{r}}-\frac{\hat{r}}{6}-1\right).
\label{col2}
\end{eqnarray}
Thus, the complete in-medium modification 
to the Coulomb term (Eqs. \ref{col1} and \ref{col2}), for the parallel alignment 
becomes
\begin{eqnarray}
 V_1({\bf r}||{\bf n}, \xi,T) &=&V_1^{(1)}({ r},\xi=0,T)+ 
V_1^{(2)} ({\bf r}||{\bf n},\xi,T) \nonumber\\
&=&-\frac{\alpha m_{{}_D}}{\hat{r}}e^{-\hat{r}}-\alpha m_{{}_D}-\xi~
\frac{\alpha m_{{}_D}}{\hat{r}}~e^{-\hat{r}}\left(2
\frac{e^{\hat r}-1}{{\hat r}^2}-\frac{2}{\hat r}-\frac{\hat r}{6}-1\right).
\label{col}
\end{eqnarray}
Next we consider the other scenario, i.e.  when $ \mathbf{r} $ is 
transverse to the direction of anisotropy, ${\bf n}$ 
where we take {\bf r} along the $z$-axis and {\bf n} lying in the $x$-$y$ plane, so
$\phi$ is the azimuthal angle and $\phi_n$ is the angle between 
$\mathbf{n}$ with $x$-axis.  Then $V_1^{(2)}({\bf r} \perp {\bf n},\xi,T)$ 
becomes
\begin{eqnarray}
V_1^{(2)}({\bf r \perp n},\xi,T)\!\!\!\!\!&=&\!\!\!\!\!-\xi
~\frac{\alpha m_{{}_D}^2}{2\pi^2}\int 
\frac{d^3\mathbf{k}~e^{i {\bf k} \cdot {\bf r}} }{(k^2+m_{{}_D}^2)^2}~
\left(\frac{2}{3}-\cos^2
\left(\phi-\phi_n\right)\sin^2\theta\right).
\end{eqnarray}
After the angular integration (exploiting the cylindrical symmetry), it 
is simplified into
\begin{eqnarray}
V_1^{(2)}({\bf r \perp n},\xi,T)=\xi V_1^{(1)}({ r},\xi=0,T)\left(\frac{1-e^{\hat{r}}}{{\hat{r}}^2}
+\frac{1}{\hat{r}} +\frac{\hat{r}}{3}+\frac{1}{2}\right)
\label{colperp}
\end{eqnarray}
Thus, the complete in-medium modification 
to the Coulomb term (Eqs. \ref{col1} and \ref{colperp}), for the transverse 
alignment becomes
\begin{eqnarray}
V_1({\bf r\perp n},\xi, T) &=& V_1^{(1)}({ r},\xi=0,T)
        + V_1^{(2)}({\bf r\perp n},\xi,T) \nonumber\\
      &=&-\frac{\alpha m_{{}_D}}{\hat{r}}e^{-\hat{r}}-\alpha m_{{}_D}
       +\xi~\frac{\alpha m_{{}_D}}{\hat{r}} e^{-\hat{r}}
       \left(\frac{e^{\hat{r}}-1}{{\hat{r}}^2}-\frac{1}{\hat{r}}
      -\frac{\hat{r}}{3} -\frac{1}{2}\right).
\label{couperp}
\end{eqnarray}
Next we calculate the in-medium-modification to the linear term, 
$V_2({\bf r},\xi,T)$ where the isotropic part (from (\ref{defn})) is given by 
\begin{equation}
V_2^{(1)}({r},\xi =0,T)=\frac{2\sigma}{m_{{}_D} \hat{r}}
e^{-\hat{r}}-\frac{2\sigma}{m_{{}_D} \hat{r}}+\frac{2\sigma}{m_{{}_D}}
\label{lin1}
\end{equation}
and the anisotropic contribution, when $ \mathbf{r} $ is parallel to the 
direction of anisotropy $\bf n$, is given by
\begin{eqnarray}
V_2^{(2)}({\bf r || n},\xi,T) &=&-\xi~\frac{4\sigma}{(2\pi)^2}\int 
\frac{d^3\mathbf{k}~e^{i \mathbf{k} \cdot \mathbf{r}}}{k^2~(k^2+m_{{}_D}^2)^2}
~\left(\frac{2}{3}-\cos^2\beta_n\right) .
\end{eqnarray}
After the angular integration, it becomes
\begin{eqnarray}
V_2^{(2)}({\bf r || n},\xi,T)
&=& -\xi \frac{4\sigma}{m_{{}_D} \hat{r}}{e^{-\hat{r}}}\left(2\frac{(1-e^{\hat r})}{\hat r^2}+
\frac{e^{\hat r}+2}{3} +\frac{2}{\hat r}+\frac{\hat {r}}{12}\right)
\label{lin2}
\end{eqnarray}
Thus, the complete in-medium modification to the linear term (Eqs. \ref{lin1} 
and \ref{lin2}), for $ \mathbf{r} \ || \mathbf{n} $ becomes
\begin{eqnarray}
V_2({\bf r|| n}, \xi, T) &=&
V_2^{(1)}({r},\xi=0,T)+ V_2^{(2)}({\bf r|| n},\xi,T)
\nonumber\\
      &=&-\frac{2\sigma}{m_{{}_D} \hat{r}}
   + \frac{2\sigma}{m_{{}_D} \hat r} e^{-\hat{r}}+\frac{2\sigma}{m_{{}_D}}\nonumber\\
     &+&\xi~\frac{4\sigma}{m_{{}_D} \hat{r}}{e^{-\hat{r}}}\left(2~
\frac{e^{\hat{r}}-1}{{\hat{r}}^2}-\frac{e^{\hat{r}}+2}{3} -\frac{2}{\hat{r}}
-\frac{\hat{r}}{12}\right)
\label{lin}
\end{eqnarray}
On the other hand, when $ \mathbf{r} $ is transverse to the direction of anisotropy,
$\mathbf{n}$, the anisotropic contribution to the linear term becomes
\begin{eqnarray}
V_2^{(2)}({\bf{r} \perp\bf{n}},\xi,T) &=&-\xi~\frac{4\sigma}{(2\pi)^2}\int 
\frac{d^3\mathbf{k}~e^{i \mathbf{k} \cdot \mathbf{r}}}{k^2~(k^2+m_{{}_D}^2)^2}
~ \left(\frac{2}{3}-\cos^2 \left(\phi-\phi_n\right)\sin^2\theta\right), 
\end{eqnarray}
which is simplified into
\begin{eqnarray}
V_2^{(2)}({\bf{r} \perp\bf{n}},\xi,T) &=&
-\xi ~\frac{4\sigma}{m_{{}_D} \hat{r}}{e^{-\hat{r}}}\left(\frac{(e^{\hat r}-1)}{\hat r^2}
+\frac{(e^{\hat r}-7)}{12}-\frac{1}{\hat r}-\frac{\hat r}{6}\right) ,
\label{linp}
\end{eqnarray}
after the angular integration. Thus, the complete in-medium modification to the linear term 
(Eqs. \ref{lin1} and \ref{linp})
for $ \mathbf{r} \perp \mathbf{n} $ becomes
\begin{eqnarray}
V_2({\bf{r}\perp\bf{n}}, \xi, T) &=&
V_2^{(1)}({r},\xi=0,T)+ V_2^{(2)}({\bf r\perp\bf{n}},\xi,T)
\nonumber\\
&=&-\frac{2\sigma}{m_{{}_D} \hat{r}}
+ \frac{2\sigma}{m_{{}_D} \hat{r}} e^{-\hat{r}}+\frac{2\sigma}{m_{{}_D}}\nonumber\\
&-&\xi ~\frac{4\sigma}{m_{{}_D}^2 r}{e^{-m_{{}_D} r}}\left(\frac{(e^{\hat r}-1)}{\hat r^2}
+\frac{(e^{\hat r}-7)}{12}-\frac{1}{\hat r}-\frac{\hat r}{6}\right).
\label{linperp}
\end{eqnarray}
\par Therefore, the full medium-modified potential, consisting of Coulomb
and linear term (Eqs. \ref{col} and \ref{lin}, respectively), for the
quark pairs aligned to the direction of anisotropy, yields as 
\begin{eqnarray}
V({\bf{r} \parallel \bf{n}},\xi,T)
&=& \left(\frac{2\sigma}{m_{{}_D}}-\alpha m_{{}_D} \right)\frac{e^{-\hat r}}{\hat{r}}
-\frac{2\sigma}{m_{{}_D} \hat{r}}
+\frac{2\sigma}{m_{{}_D}}- \alpha m_{{}_D} \nonumber\\
&+& \xi~\left[ \frac{4\sigma}{m_{{}_D} \hat{r}}~e^{-\hat r} \left( {} 2~\frac{e^{\hat r}-1}
{\hat r^2}-\frac{e^{\hat r}+2}{3} -\frac{2}{\hat r}-\frac{\hat r}{12} \right) \right. \nonumber\\
&-&\left. \frac{\alpha m_{{}_D} }{\hat{r}} e^{-\hat r}\left( {} 2~\frac{(e^{\hat r}-1)}{\hat r^2}
-\frac{2}{\hat r}-\frac{\hat r}{6} -1\right) \right] \nonumber\\
&\equiv & V_{\rm{iso}} (r,T)+ V_{\rm{aniso}}^\parallel (r,\xi,T).
\label{fullpotpara}
\end{eqnarray}
In the short distance limit ($r \ll 1/m_{{}_D}$), the potential 
reduces to the vacuum potential (Cornell) for $\xi=0$, {\it i.e.} 
$Q\bar Q$ pairs are not affected by 
the medium. On the other hand, in the long-distance limit
($r\gg 1/m_{{}_D}$) (where the screening occurs), we can 
neglect the Yukawa term and for large values of temperatures, the product
$\alpha m_{{}_D}$ will be much greater than $2\sigma/m_{{}_D}$. Thus
the potential is simplified into the following form:
\begin{eqnarray}
V({\mathbf r} \parallel {\mathbf n},\xi,T) &\stackrel{\hat{r}\gg 1}{\simeq} & -\frac{2\sigma}{m_{{}_D}^{2}r} - 
\alpha m_{{}_D} -\frac{4\xi}{6}~\left(\frac{2\sigma}{m_{{}_D}^{2}r} \right) \nonumber\\
&\stackrel{\hat{r}\gg 1}{\equiv} & V_{\rm{iso}} (\hat{r} \gg 1,T)+ V_{\rm{aniso}}^\parallel
(\hat{r} \gg 1,\xi,T),
\label{paral}
\end{eqnarray}
which clearly shows that the potential for $Q \bar Q$ pairs aligned in the 
direction of anisotropy gets screened less, {\em i.e.} becomes stronger 
compared to isotropic medium (\ref{lrp}).

\noindent On the other hand, when the quark pairs are aligned transverse to the 
direction of anisotropy ($ \mathbf{r}  \perp \mathbf{n} $), 
the medium modification to the Coulombic and linear terms together 
(Eqs. \ref{couperp} and \ref{linperp}, respectively) gives rise to 
the following form:
\begin{eqnarray}
V({\bf r\perp {\bf n}},\xi,T)&=& 
\left(\frac{2\sigma}{m_{{}_D} }-\alpha m_{{}_D} \right) 
\frac{e^{-\hat r}}{\hat{r}}-\frac{2\sigma}{m_{{}_D}\hat{r}}  
+\frac{2\sigma}{m_{{}_D}}- \alpha m_{{}_D} \nonumber\\
&-&\xi~\left[ \frac{4\sigma}{m_{{}_D} \hat{r}}~{e^{-\hat{r}}} 
\left(\frac{(e^{\hat r}-1)}{\hat r^2}
+\frac{(e^{\hat r}-7)}{12}-\frac{1}{\hat r}-\frac{\hat r}{6}\right) \right.
\nonumber\\
&-& \left. \frac{\alpha m_{{}_D}}{\hat{r}} e^{-\hat{r}}
\left(\frac{e^{\hat r}-1}{\hat r^2}-\frac{1}{\hat r}-\frac{1}{2}
-\frac{\hat r}{3}\right) \right] \nonumber\\
&\equiv & V_{\rm{iso}} (r,T)+ V_{\rm{aniso}}^\perp (r,\xi,T),
\label{fullpotperp}
\end{eqnarray}
Similarly, $Q \bar Q$ pair does not see the medium, in the short-distance 
limit whereas in the long-distance limit, the 
potential is simplified into a Coulombic form with
a dynamically screened color charge ($2\sigma/m_D^2$):
\begin{eqnarray}
V({\bf r} \perp {\bf n},\xi,T)&\stackrel{\hat{r}\gg1}{\simeq}&-\frac{2\sigma}{m_{D}^{2}r}
-\alpha m_{D}-\frac{\xi}{6} \left(\frac{2\sigma}{m_{D}^{2}r}\right)\nonumber\\
&\stackrel{\hat{r}\gg 1}{\equiv} & V_{\rm{iso}} (\hat{r}\gg 1,T)+ V_{\rm{aniso}}^\perp 
(\hat{r}\gg 1,\xi,T),
\label{perp} 
\end{eqnarray}
which again shows that the potential for the transverse alignment is still
stronger than in isotropic medium but less stronger than the former.

\par Until now we have demonstrated the effects of momentum anisotropy on
the heavy-quark interaction for the special cases {\em viz} the inter-particle 
separation (r) may be parallel or perpendicular to the direction of 
anisotropy (n). We would now like now to derive a potential which depends 
on both the inter-particle separation (r) and the
angle ($\theta_n$) between ${\bf r}$ and $n$, explicitly.

Let us assume that ${\bf r}$  is parallel to the z component of ${\bf k}$
and the direction of anisotropy $ {\bf n} $ lies in the $ y$-$z$
plane (cylindrical symmetry) in the momentum space.
We may further assume that given the weak anisotropy, the potential in 
the anisotropic medium represents a perturbation to the central potential 
in the isotropic medium as : 
\begin{eqnarray}
V(r,\theta_n, T)&=&V(r,T)+ V_{\rm{tensor}}(r,\theta_n,T)\\
&=& V_{\rm{iso}}(r,T)+\xi F(r,\theta_n,T).
\label{potgen}
\end{eqnarray}
where the tensorial part $V_{\rm{tensor}}(r,\theta_n,T)$ represents a small
perturbation to the central one $V(r;T)$ and $\xi$ is the strength of 
non-central component of the potential. The function $F(r,\theta_n,T)$ can 
be expanded as
\begin{equation}
F(r,\theta_n, T)=f_0(r,T)+f_1(r,T)\cos 2\theta_n~.
\end{equation}
The potentials for the angles $\theta_n=0$ and $\theta_n=\pi/2$ help us to 
determine the functions $ f_0(r,T) $ 
and $f_1(r,T)$ in terms\footnote{The variable
$\hat{r}$ should not be confused with the usual notation of unit vector in
the coordinate system.} of $\hat{r}$ (=$rm_D$) as
\begin{eqnarray}
f_0(\hat{r},T)&=&\frac{2 \sigma} {m_{D}}\frac{e^{-\hat{r}}}{\hat{r}} 
\left(\frac{e^{\hat{r}}-1}{\hat{r}^2}-\frac{5 e^{\hat{r}}}{12}-
\frac{1}{\hat{r}}+\frac{\hat{r}}{12} -\frac{1}{12}\right)\nonumber\\
&-&\frac{\alpha m_{D}} {2}\frac{e^{-\hat{r}}}{\hat{r}}
\left(\frac{e^{\hat{r}}-1}{\hat{r}^2}-\frac{1} {\hat{r}}+
\frac{\hat{r}}{6}-\frac{1}{2}\right)
\end{eqnarray}
and 
\begin{eqnarray}
f_1(\hat{r},T)&=&\frac{2 \sigma} {m_{D}}\frac{e^{-\hat{r}}}{\hat{r}}\left(
3 \frac{e^{\hat{r}}-1}{\hat{r}^2}-\frac{e^{\hat{r}}}{4}-
\frac{3}{\hat{r}}-\frac{\hat{r}}{4}-\frac{5}{4}\right)\nonumber\\
&-&\frac{\alpha m_{D}}{2} \frac{e^{-\hat{r}}}{\hat{r}}\left(
3 \frac{e^{\hat{r}}-1}{\hat{r}^2}-\frac{3}{\hat{r}}-\frac{\hat{r}}{2}-
\frac{3}{2}\right)
\end{eqnarray}
So after substituting $f_0$ and $f_1$ into the Eq.(\ref{potgen}), we obtain
the complete angular dependence of the potential in 
the limit of weak anisotropy ($\xi \ll 1$) :
\begin{eqnarray}
V(r,\theta_n,T) &=& \left(\frac{2\sigma}{m_{{}_D} }-\alpha m_{{}_D} \right) 
\frac{e^{-\hat r}}{\hat{r}}-\frac{2\sigma}{m_{{}_D}\hat{r}}  
+\frac{2\sigma}{m_{{}_D}}- \alpha m_{{}_D} \nonumber\\
&+&  \xi \left( \frac{2 \sigma} {m_{D}}\frac{e^{-\hat{r}}}{\hat{r}} 
\left[\frac{e^{\hat{r}}-1}{\hat{r}^2}-\frac{5 e^{\hat{r}}}{12}-
\frac{1}{\hat{r}}+\frac{\hat{r}}{12} -\frac{1}{12}\right] \right. \nonumber\\
&-& \left. \frac{\alpha m_{D}} {2}\frac{e^{-\hat{r}}}{\hat{r}}
\left[\frac{e^{\hat{r}}-1}{\hat{r}^2}-\frac{1} {\hat{r}}+
\frac{\hat{r}}{6}-\frac{1}{2}\right] \right. \nonumber\\
&+ & \left( \left. \frac{2 \sigma} {m_{D}}\frac{e^{-\hat{r}}}{\hat{r}}\left[
3 \frac{e^{\hat{r}}-1}{\hat{r}^2}-\frac{e^{\hat{r}}}{4}-
\frac{3}{\hat{r}}-\frac{\hat{r}}{4}-\frac{5}{4}\right] \right.\right.\nonumber\\
&-&\frac{\alpha m_{D}}{2} \frac{e^{-\hat{r}}}{\hat{r}}\left[ \left.\left.
3 \frac{e^{\hat{r}}-1}{\hat{r}^2}-\frac{3}{\hat{r}}-\frac{\hat{r}}{2}-
\frac{3}{2}\right] \right) \cos~2 \theta_n \right) \nonumber\\
&=& V(r,T) + V_{\rm{tensor}}(r,\theta_n,T) 
\label{fullp}
\end{eqnarray}
Thus the anisotropy in the momentum space introduces an angular 
($\theta_n$) dependence, in addition to the inter-particle separation ($r$),
to the potential in the coordinate space which was earlier only $r$-dependent 
in the isotropic medium. We can now identify the $\xi$-independent 
term with $V(r,T)$ in (50) which depends only on the separation ($r$) distance 
and the $\xi$-dependent term with the tensorial component 
$V_{\rm{tensor}}(r,\theta_n,T)$ in (50) which depends on both 
$r$ and $\theta_n$. Thus the full potential in an anisotropic 
medium, $V(r, \theta_n,T)$ needs to be solved by the three-dimensional 
Schr\"odinger equation. Since we are restricted in the small $\xi$ limit, so 
the tensorial component $V_{\rm{tensor}}(r,\theta_n,T)$ is much smaller than
the (isotropic) central component $V(r,T)$ and hence may be treated as the 
perturbation by a first-order perturbation theory in quantum mechanics.
However, the isotropic component may be solved numerically by the 
one-dimensional (radial) Schr\"odinger equation.

The heavy quark interaction at short and intermediate distances
($rm_D \le1$) are important for the understanding of in-medium
modification of heavy quark bound states and the large distance
property ($rm_D>1$) helps to understand the bulk properties of
QGP phase which also affect the in-medium properties
of the quarkonium states. So we wish to 
to see how the potential in anisotropic medium behaves in these
(short, intermediate and long) limiting cases.
In the short-distance limit, the vacuum contribution 
dominates over the medium contribution and this is exactly happens 
here 
\begin{eqnarray}
V(r,\theta_n,T)\stackrel{\hat{r}\ll 1}{\simeq} \sigma r - \frac{\alpha}{r}
\label{vprime}
\end{eqnarray}                                        
for $\xi=0$. On the other hand, in the long-distance limit ($\hat{r}\gg 1 $), the 
potential is reduced to a long-range Coulombic interaction after
identifying the factor $2\sigma/m_D^2$ with the coupling ($g_s^2$)
of the interaction
\begin{eqnarray}
\label{largp}
V(r,\theta_n,T) &\stackrel{\hat{r}\gg1}{\simeq}& -\frac{2\sigma}{m^2_{{}_D}r}-\alpha m_{{}_D}
-\frac{5\xi}{12}~\frac{2\sigma}{m^2_{{}_D}r} \left(1+\frac{3}{5}\cos 2\theta_n \right)\nonumber\\
&\equiv & V_{\rm{iso}} (\hat{r} \gg 1,T)+ V_{\rm{tensor}} (\hat{r} \gg 1~.
\theta_n,T)~.
\end{eqnarray}
Since the resulting potential is Coulombic plus a
subleading anisotropic contribution, it then has to satisfy
the condition: $a_0 m_D\gg 1$, where $a_0$ is the Bohr radius and $m_D$ is
the Debye mass. Since the Bohr radius $a_0$ is proportional to 
$m_D^2/(m_Q \sigma)$,
the above condition for the long-distance limit implies that 
$m_D^3/(m_Q \sigma)$ 
should be greater than 1. Thus this inequality results in a condition on the
Debye mass and hence on the temperature. It is seen that the above condition 
is satisfied for the temperatures above the critical temperature ($>T_c$)
for the charmonium states and above 1.6$T_c$ 
for the bottomonium states. The temperature ranges ($T_c$ and 
1.6$T_c$) for $c \bar c$ and $b\bar b$ states above which 
the effective potential looks Coulombic are
smaller than their respective dissociation temperatures and thus seems 
justified to approximate the potential in the long-distance limit. In the intermediate distance ($rm_D \simeq 1$) scale, the 
interaction becomes complicated and thus the potential
does not look simpler in contrast to the asymptotic limits, so
this limit needs to be dealt numerically with the full potential in
a Schr\"odinger equation.

\par We have thus noticed overall that in the short 
distance limit, the potential have not been affected in the isotropic limit.
On the contrary, in the long-distance limit, the momentum anisotropy 
transpires an angular dependence in the potential 
and gives rise a characteristic angular ($\theta_n$) dependence 
between the relative 
separation ($\mathbf{r}$) and the direction of anisotropy ($\mathbf{n}$). 
As a corollary, the quark pairs aligned 
along the direction of anisotropy feel more attraction than the transverse 
alignment because the inter-quark potential along
the direction of anisotropy is screened less than the transverse alignment.
However, the potential in the anisotropic medium is always stronger than in 
isotropic medium.

To see the effects of anisotropy, we have shown the potentials 
for $Q \bar Q$ pairs in an anisotropic medium in Figures 1 and 2,
for $\theta_n=0$ (parallel) and $\theta_n=\pi/2$
(perpendicular), respectively. The immediate observation
common to all figures is that the inter-quark potential in anisotropic
medium is always more attractive than in isotropic medium. 
This can be understood physically: In the small 
anisotropic limit, the anisotropic distribution function may be 
obtained from an isotropic distribution $f_{iso} (|\mathbf{k}|)$
by removing particles with a large momentum component along
$\mathbf{n}$ {\it i.e.}
$f_{iso}(\sqrt{\mathbf{k}^{2} + \xi(\mathbf{k}.\mathbf{n})^{2}}$.
This transpires in the reduction of the number of partons (around
a static test heavy quark) than in isotropic medium {\it i.e.} $n_{\rm{aniso}} (\xi)= 
n_{\rm{iso}}/\sqrt{1+\xi}$. Therefore, the (effective) Debye mass 
always becomes smaller and results in less screening of the potential than 
in isotropic medium.
\begin{figure}
\begin{subfigure}{
\includegraphics[width=7.9cm,height=8.5cm]{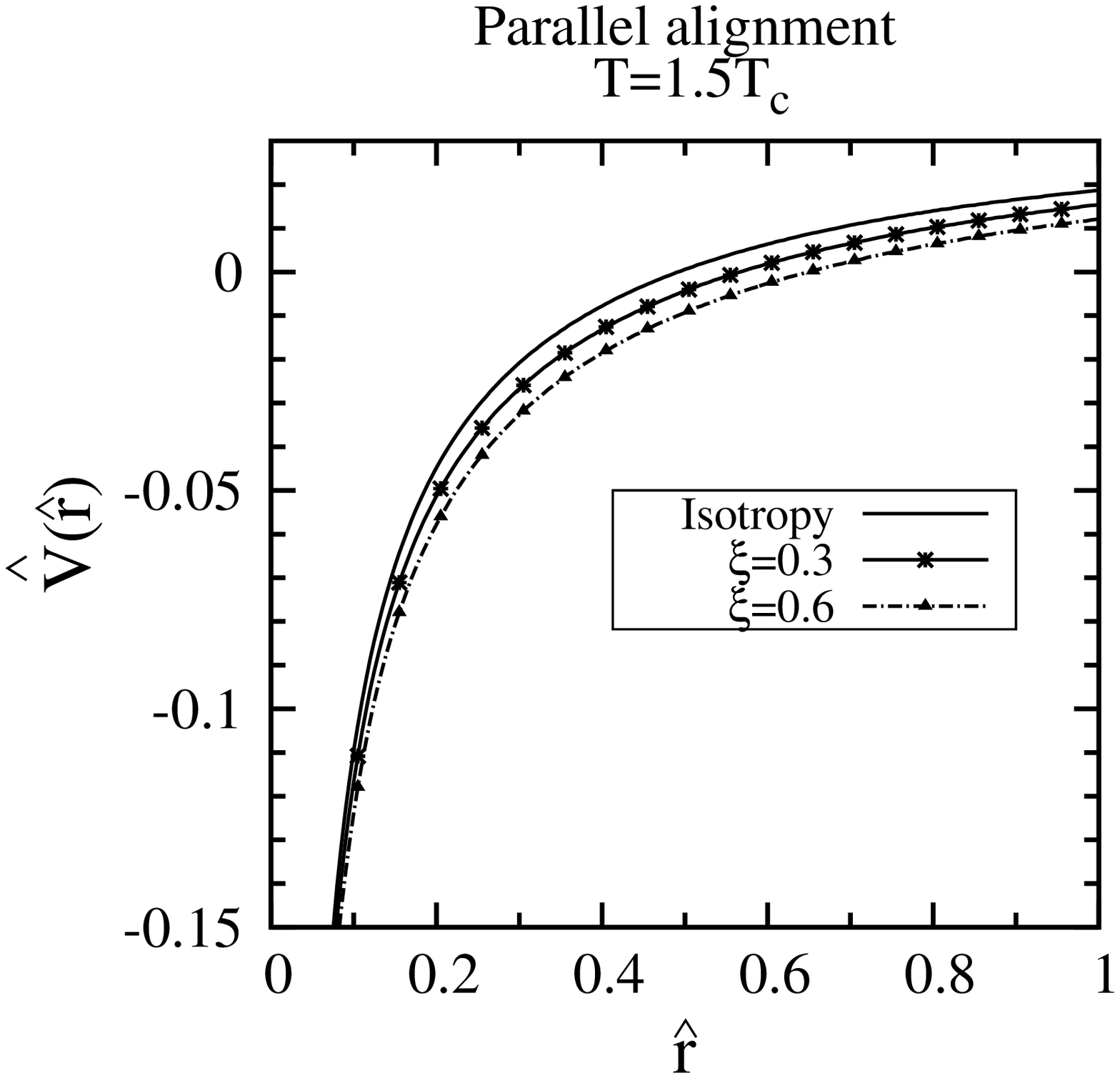}
\label{para_plot_a}}
\end{subfigure}
 \hspace{-5mm}
\begin{subfigure}{
 \includegraphics[width=7.9cm,height=8.5cm]{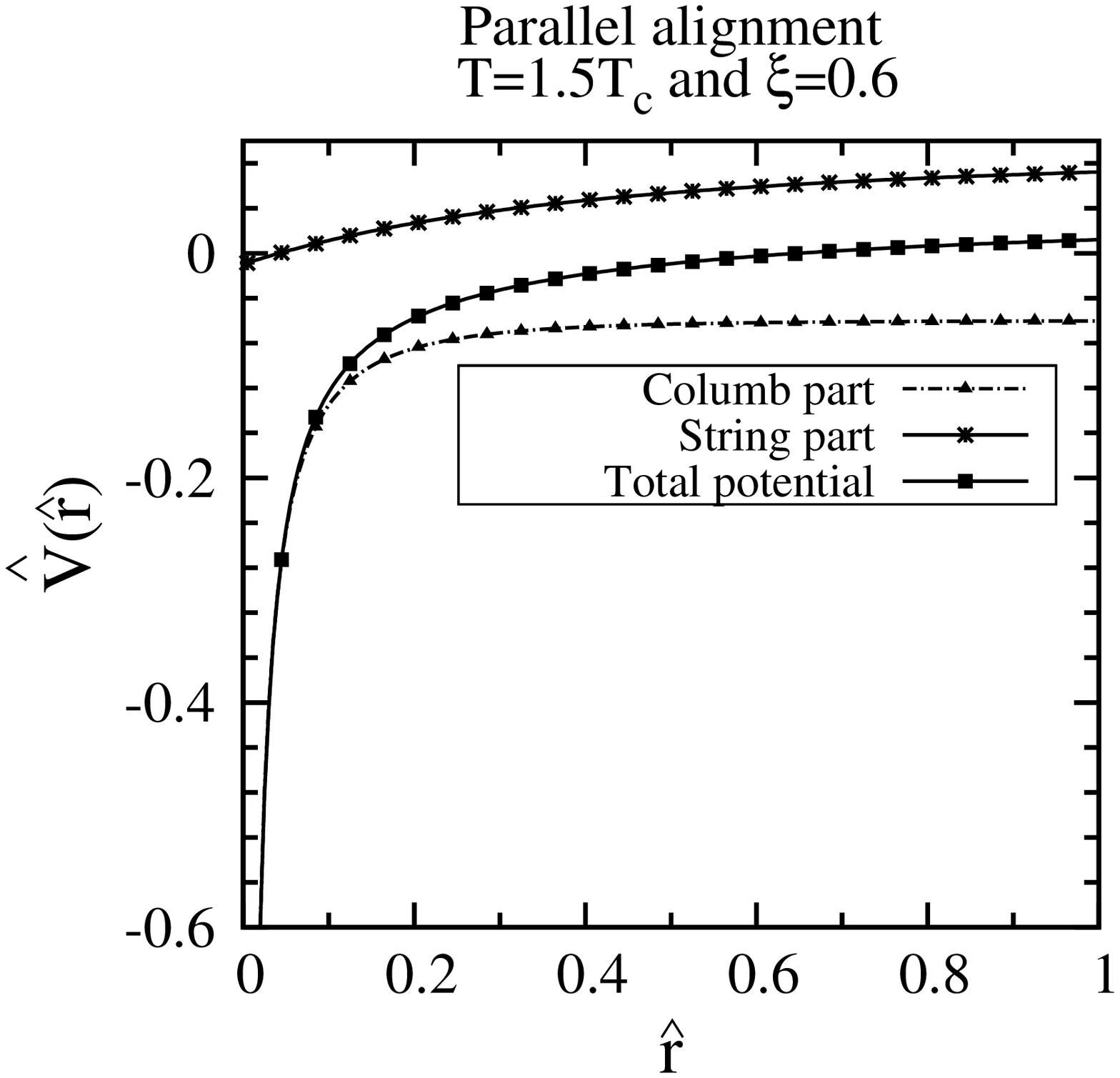}
\label{para_plot_b}}
\end{subfigure}
\caption{\footnotesize The left panel represents the potential 
divided by $(g^2 C_F m_{{}_D})$ and the right panel represents the 
contribution of Coulomb, string  and both together as a function of 
$\hat{r}$ (= $rm_{{}_D} $) for quark pairs parallel to the 
direction of anisotropy, $\mathbf{n}$.} 
\end{figure}

\begin{figure}
\vspace{0mm}
\begin{subfigure}{
\includegraphics[width=7.9cm,height=8.5cm]{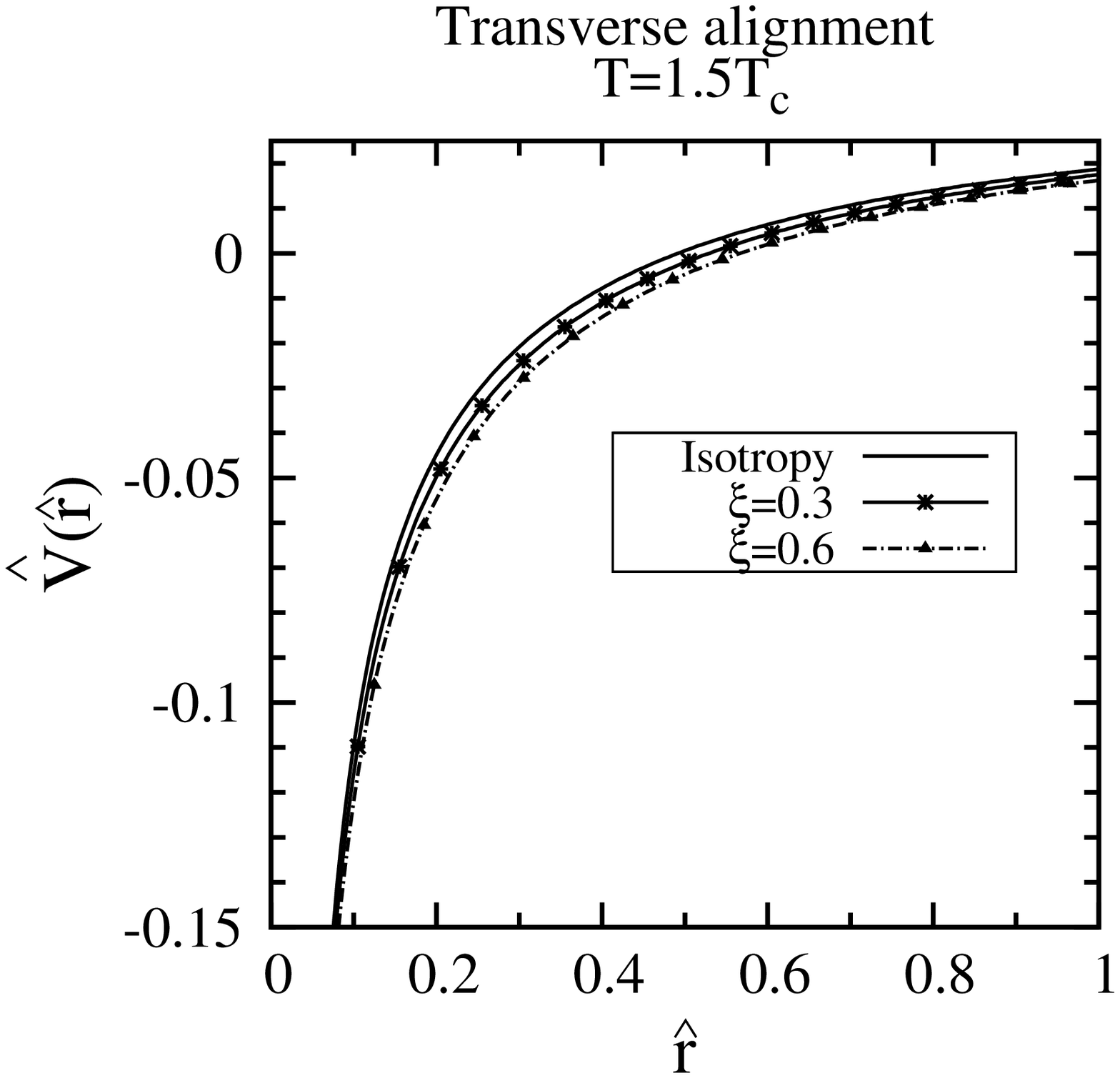} 
\label{perp_plot_a}}
\end{subfigure}
\hspace{-5mm}
\begin{subfigure}{
\includegraphics[width=7.9cm,height=8.5cm]{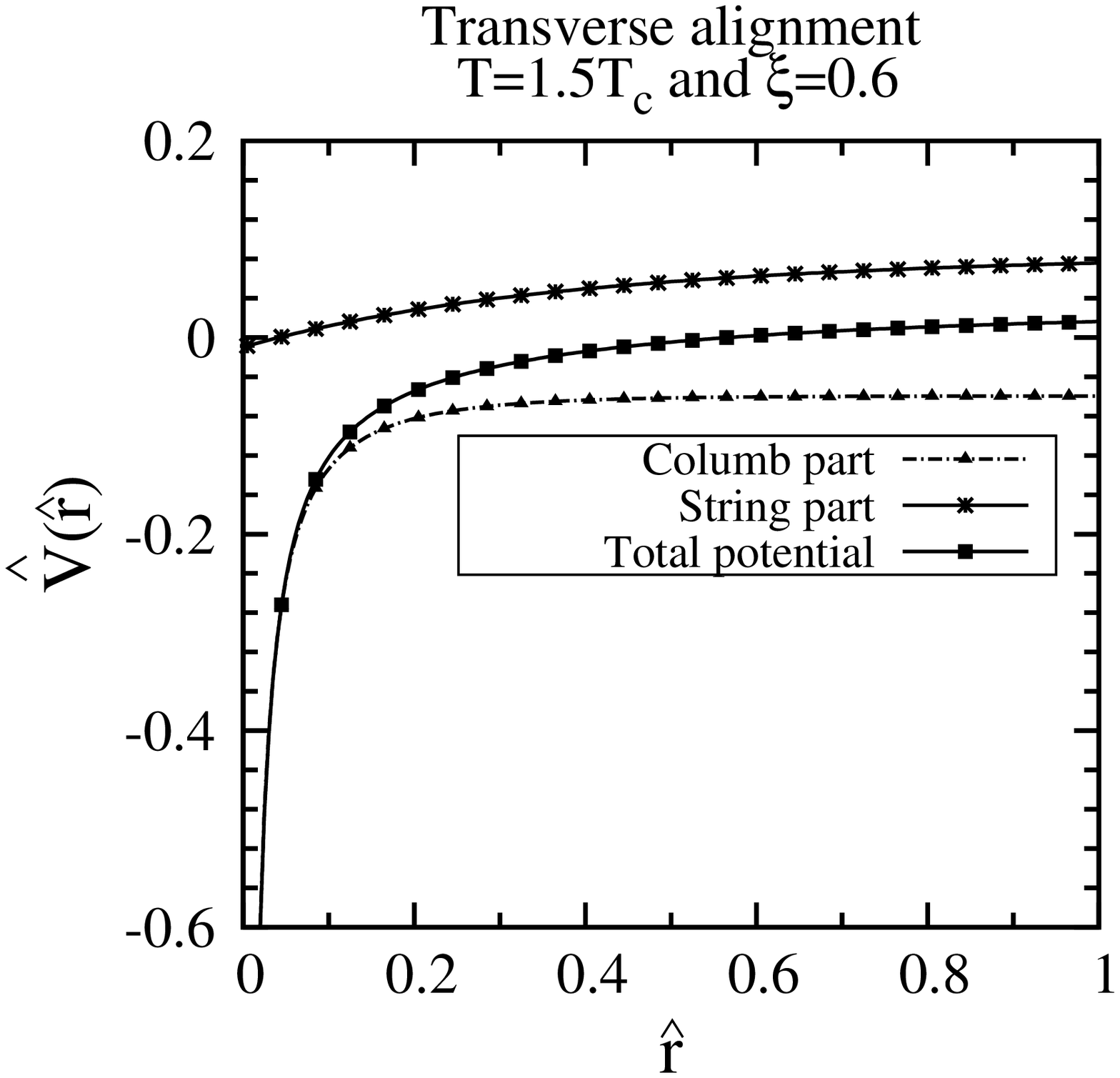}
\label{perp_plot_b}}
\end{subfigure}
\caption{\footnotesize The notations are same as in Figure 1
but for quark pairs perpendicular to the 
direction of anisotropy, $\mathbf{n}$.} 
\vspace{0mm}
\end{figure}
The second observation is that the quark pairs aligned along ($\theta_n=0$) the
direction of anisotropy are stronger than aligned perpendicular 
($\theta_n=\pi/2$) to the direction of anisotropy because for the parallel alignment, the component of momentum to be removed is higher
than the transverse alignment so the distribution function 
for the parallel alignment case contributing to the Debye mass is
smaller than the transverse alignment.
Hence the potential for parallel case will be screened less compared to 
the transverse case. However the difference between the two scenario 
will be not much different
because the contributions to the Debye mass from the partons having higher 
momenta are very small.
 
To understand the effect of linear term on the medium modified 
potential quantitatively, in addition to the Coulomb term, we have 
plotted separately the medium modifications to the linear term, 
the Coulomb term and their sum in the right panels of 
Figure 1 and 2, for parallel and transverse case, respectively.
Medium modification to the Cornell potential contains two parts: one is due to
the medium modifications of linear term ($\sigma r$) and the other one is due 
to the medium modifications of
Coulomb term. As usually done in the literature, medium 
modification to the linear term does not arise because
the string tension was assumed to be 
zero~\cite{mocsyprd,Satz,shro,Alb05} 
at or beyond deconfinement temperature~\cite{Lusch}.
Since string tension is found to be nonzero at $T_c$ rather it 
approaches zero much beyond $T_c$~\cite{string1,string2,string3} and hence
the medium modification to the linear term may be
non-zero contribution to the potential even at temperatures beyond $T_c$, although it is very small. 
In isotropic medium, medium modification to the linear term
remains  positive up to 2-3 $T_c$, making the potential less attractive 
compared to $T=0$. On contrast, in anisotropic case medium 
modification to the linear term becomes negative
and the overall full potential becomes more attractive.

As mentioned earlier we used the same screening scale for both the linear
and Coulombic terms in our calculation which does not look plausible.
It would thus be interesting to see the effects of different scales for the 
Coulomb and linear pieces of the T=0 potential~\cite{megiasind,megiasprd}.
To illustrate it graphically, we have compared our results with their results
(in Figure 3) for the isotropic case. The difference in the large
distance limit arises due to the difference in the potential
at infinity (-$\sigma/m_D$) so the 
potential in Ref.\cite{megiasprd} is more attractive than our potential.
\begin{figure}
\vspace{0mm}
\includegraphics[width=7.9cm,height=8.5cm]{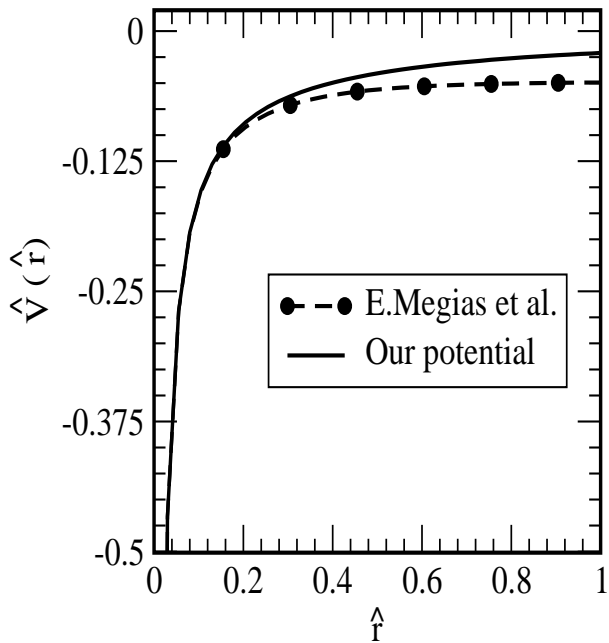} 
\caption{\footnotesize The dotted-circle represents the results from 
Megias et al. \cite{megiasprd} where
different scales was used for linear and Coulomb terms separately whereas
the solid line represents our work.}
\vspace{0mm}
\end{figure}

\section{Properties of Quarkonium in an Anisotropic Medium}\label{prop_q}
\subsection{Binding energy}
To understand the in-medium properties of the quarkonium states,
we need to model the heavy quark potential as a function of
temperature and solve the resulting Schr\"{o}dinger equation.
The potential thus obtained in anisotropic medium (\ref{fullp}), 
in contrast to the (spherically symmetric) potential in isotropic medium,
is non-spherical and so one cannot simply obtain the energy eigen values 
by solving the radial part of the Schr\"{o}dinger equation only because 
the radial part is no longer sufficient due to 
the angular dependence in the potential. Other way to understand
is that because of the anisotropic
screening scale, the wave functions are no longer radially symmetric for
$\xi \ne 0$.  So one has to solve the potential in anisotropic medium
through the Schr\"odinger equation in three dimension. 
However, we have seen in the potential (55) that in the small $\xi$-limit, 
the spherically non-symmetric component
$V_{\rm{tensor}}(r,\theta_n,T)$ is much smaller in comparison to spherically 
symmetric (isotropic) component $V(r,T)$ and thus
can be treated as perturbation. This can be understood physically:
The tensorial (non-sphericity) nature
of the potential in the co-ordinate space is arisen due to anisotropy 
in the momentum space. However, we are restricted to a plasma which
is very much close to equilibrium because
by the time quarkonium states are formed in the plasma
around (1-2)$T_c$, the plasma becomes almost 
isotropized. Thus  this weak (momentum) anisotropy
($\xi \ll 1$) transpires feeble angular dependence in the potential
so the potential will be spherically abundant
with a tiny non-spherical component. So we could 
treat the anisotropic component through the perturbation theory in quantum 
mechanics and the isotropic part should be
handled numerically by the one-dimensional radial
Schr\"{o}dinger equation.

There are some numerical methods to solve the Schr\"odinger equation
either in partial differential form (time-dependent) or eigen value form
(time-independent/stationary) by the
finite difference time domain method (FDTD) or matrix method, respectively.
However, we choose the matrix method to solve the
stationary Schr\"odinger equation with the isotropic part of the potential
(55) in anisotropic medium. In this method, the Schr\"odinger equation
can be cast in a matrix form through a discrete basis, instead
of the continuous real-space position basis spanned by the states
$|\overrightarrow{x}\rangle$. Here the confining potential V is subdivided
into N discrete wells with potentials $V_{1},V_{2},...,V_{N+2}$ such that
for $i^{\rm{th}}$ boundary potential, $V=V_{i}$ for $x_{i-1} < x < x_{i};~i=2, 3,..
.,(N+1)$. Therefore for the existence of a bound state, there
must be exponentially decaying wave function
in the region $x > x_{N+1}$ as $x \rightarrow  \infty $ and
has the form:
\begin{equation}
\Psi_{N+2}(x)=P_{{}_E} \exp[-\gamma_{{}_{N+2}}(x-x_{N+1})]+ 
Q_{{}_E} \exp [\gamma_{{}_{N+2}}(x-x_{N+1})] , 
\end{equation}
where, $P_{{}_E}= \frac{1}{2}(A_{N+2}- B_{N+2})$,
$Q_{{}_E}= \frac{1}{2}(A_{N+2}+ B_{N+2}) $ and,
$ \gamma_{{}_{N+2}} = \sqrt{2 \mu(V_{N+2}-E)}$. The eigenvalues
can be obtained by identifying the zeros of $Q_{E} $.

\begin{figure*}
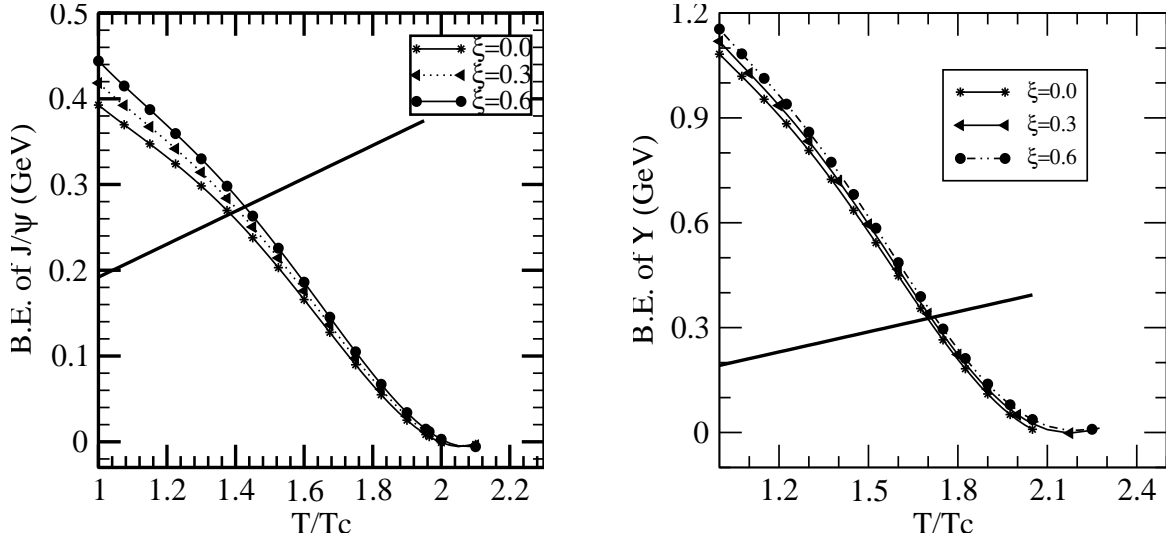

\begin{subfigure}{
\includegraphics[trim = 1mm 1mm 1mm 1mm, clip,height=7cm,width=7cm]{charm.eps}
}
\end{subfigure}
\hspace{5mm}
\begin{subfigure}{
\includegraphics[trim = 1mm 1mm 1mm 1mm, clip,height=7cm,width=7cm]{bot.eps}
}
\end{subfigure}
\vspace{0mm}
\caption{\footnotesize Variation of $J/\psi$ and $\Upsilon$ binding 
energy (in $GeV$) with the temperature in an anisotropic hot QCD medium.}
\label{bind_jsi}
\end{figure*}

Therefore, the corrected energy eigen value comes
from the solution of Schr\"odinger equation
of the isotropic component $V_{\rm{iso}}(r,T)$, using the 
abovementioned matrix method
plus the first-order perturbation due to the anisotropic 
component $V_{\rm{aniso}}(r,\theta;\xi,T)$ (55) through
the quantum mechanical perturbation theory.
The variations of the binding energies with the 
temperature are shown in figure~(\ref{bind_jsi})
for $J/\psi$ and $\Upsilon$ 
for different values of anisotropy parameter $\xi$,
to see the effect of anisotropy on the binding energies compared to
the isotropic case.  

There are mainly two observations: First, as the anisotropy increases, the binding of 
$Q \bar Q$ pairs get
stronger with respect to their isotropic counterpart because 
the potential becomes deeper with the increase of anisotropy due to
weaker screening. It seems 
that the (effective) Debye mass $m_D(\xi,T)$ in an anisotropic
medium is always smaller than in an isotropic medium. As a result
the screening of the Coulomb and string contribution are 
less accentuated and hence the quarkonium states 
become more stronger than in an isotropic medium.
However, the effects of anisotropy on the 
excited states are not 
so pronounced compared to the ground states
because they are generically weakly bound.
Secondly, there is
a strong decreasing trend with the temperature.
This is due to the fact that the screening becomes always stronger with the increase of
temperature, so the potential becomes weaker compared to $T=0$ and
results in early dissolution of quarkonia in the medium.
Our results on the temperature dependence of the binding energies show
an agreement with the similar variations in other
calculations~\cite{Dumitru09}.

In our calculation, we use the Debye mass 
($m_{{}_D}^{\rm L}=1.4 m^{\rm LO}_{{}_D}$) obtained by fitting
the (color-singlet) free energy in lattice QCD~\cite{mocsyprl} 
where both one and two-loop expression 
~\cite{shro,shaung,ijmp} for coupling have been used to explore the
effects of running coupling on the dissociation process.

Thus the study of the temperature dependence of the binding 
energies are poised
to provide a wealth of information about the dissociation
pattern of quarkonium states in an anisotropic thermal medium that can be
used to determine the dissociation temperatures of different states
in the next Section.

\subsection{Dissociation temperatures for heavy quarkonia}\label{diss}
Dissociation of a two-body bound state in an thermal medium can be 
understood qualitatively: When the binding energy of a resonance state
drops below the mean thermal energy of a parton, 
the state becomes feebly bound. The thermal
fluctuation then can easily dissociate by exciting
them into the continuum.
The spectral function technique in potential models defines the dissociation
temperature as the temperature above which the quarkonium spectral
function shows no resonance-like structures but the widths shown 
in spectral functions from current potential model calculations
are not physical. The broadening of states with the increase in 
temperature is not included in any of these models. 
In Ref.\cite{mocsyprl}, the authors
argued that one need not to reach the binding energy ($E_{\rm{bin}}$) to
be zero for the dissociation rather a weaker condition $E_{\rm{bin}}<T$
causes a state weakly bound. In fact, when $E_{\rm{bin}} \simeq T$, the
resonances have been broadened due to direct thermal activation, 
so the dissociation of the bound states may be expected to occur
roughly around $E_{\rm{bin}} \simeq T$.

Using the binding energies calculated earlier in Sec.3.1, 
the dissociation temperatures ($T_D$) (shown in Table 1)
are found minimum for the isotropic case and increase
with the increase of anisotropy ($\xi>0$) {\em viz.} $J/\psi$ 
is dissociated at $1.38~T_c$ in an isotropic medium while in an 
anisotropic medium with the anisotropies $\xi$=0.3 and 0.6, they will 
survive higher 
temperatures, $1.41~T_c$ and $1.43~T_c$, respectively. Similarly 
the dissociation temperatures of 
$\Upsilon$ for $\xi$=0.3 and 0.6 are $1.71~T_c$
and $1.72~T_c$, respectively, corresponding to the value 
($1.70~T_c$) in an isotropic medium. 

\begin{table}
\centering
\begin{tabular}{|c|c|c|c|}
\hline
State &$\xi=0.0$ & $\xi=0.3$ & $\xi=0.6$\\
\hline\hline
$\jpsi$ & 1.38& 1.41& 1.43  \\
\hline
$\Upsilon$ & 1.70 & 1.71 & 1.72 \\
\hline
\end{tabular} 
\caption{\footnotesize Dissociation temperatures ($T_D$)
for the quarkonium states with one-loop QCD coupling}
\label{tdpara1l}
\end{table}
\par Finally we wish to explore the effects of perturbative as well as
non-perturbative contributions on the dissociation of quarkonia states
qualitatively in terms of the debye mass.
Instead of lattice Debye mass ($m_D^L$), if we use the leading-order
Debye mass ($m_D^{\rm{LO}}<m_D^L$), the screening of the potential 
will be much smaller and hence the binding energies ($1/m_{{}_D}^4$) 
will be enhanced substantially
and results in the increase of the dissociation temperatures.
On the other hand, if we include the non-perturbative corrections 
of order $O(g^2T)$ and $O(g^3T)$ to the leading-order Debye mass~\cite{kajantie},
the dissociation temperatures become unrealistically small
which looks unfeasible. Thus, this study
provides us a handle to decipher the extent up to which and how much
non-perturbative effects should be incorporated into the Debye mass.

\section{Conclusions and Outlook}\label{conclu}
In conclusion, we have studied the dissociation of quarkonia 
by correcting the full Cornell potential with a dielectric function 
embodying the effects of an weakly anisotropic medium 
where the in-medium modification causes less screening of the interaction
and hence the potential gets stronger than in an isotropic medium. 
Anisotropy further introduces a characteristic
angular ($\theta$) dependence to the potential in the coordinate space, in addition to the 
inter-particle separation $r$, making it spherically non-symmetric 
and needs to be solved numerically by the three-dimensional
Schr\"odinger equation. However, in the small $\xi$-limit, the spherically 
non-symmetric component is much smaller in comparison to spherically 
symmetric component and can be treated in a perturbation theory and
the symmetric (isotropic) component is solved numerically by
one-dimensional radial Schr\"{o}dinger equation.
So the corrected binding energy is obtained by the direction-independent shift 
due to the spherically non-symmetric component to the eigen values of 
spherically-symmetric potential.

We have observed that the quarkonia states are always more bound 
and as a consequence, they survive
higher temperature compared to the isotropic medium. 
Our results are found relatively higher compared to similar 
calculation~\cite{Dumitru08}, which
may be due to the absence of three-dimensional medium modification 
of the linear term in their calculation. In fact, one-dimensional Fourier
transform of the Cornell potential yields the similar form
used in the lattice QCD in which one-dimensional color flux tube structure
was assumed~\cite{dixit}.
However, at finite temperature that may not be
the case since the flux tube structure may expand in more
dimensions~\cite{Satz}.
Therefore, it would be better to consider the three-dimensional form of the medium
modified Cornell potential which has been done exactly in the present work.

In brief, $\jpsi$ is found to be dissociated at 1.38 $T_c$ and $1.43~T_c $ 
for $\xi$=0 and 0.6, respectively
whereas the corresponding temperatures for the $\Upsilon$ state are
$1.70~T_c$ and $1.72~T_c$.
Moreover we explore the effects of perturbative as well as
non-perturbative effects on the dissociation process qualitatively.
{\em For example}, the perturbative result of Debye mass 
gives much higher values of dissociation temperatures 
whereas the inclusion of non-perturbative corrections to it
gives unrealistically smaller values. It may be important to note that in 
the weakly-coupled regime, the effects of (non-perturbative) terms {\em viz.}
$g^2T$, 
$g^3T$ etc. may be checked separately but in the strong-coupling regime, this 
may not be possible because they are no longer uncoupled.
These findings envisage a basic question about the nature of dissociation 
of quarkonium in an anisotropic hot QCD medium.

Apart from the uncertainty of the correct form of the potential,
there is an arbitrariness in the definition of dissociation temperature. So,
in future, we wish to investigate the dissociation 
through the decay width of quarkonium bound states calculated from the
imaginary part of the potential because it is now well understood that potential
in thermal medium has always an imaginary component~\cite{Laine2,nora11}. 

\noindent {\bf Acknowledgments:}
We are thankful for some financial assistance from CSIR project (CSR-656-PHY),
Government of India. We also thank M G Mustafa for some suggestion.

\end{document}